**Observation of Antichiral Hinge States in a Three-dimensional Gyromagnetic Photonic Crystal**

Ziyao Wang,[1,*], Tianzhi Xia,[2,*,†] , Han-Rong Xia[1,†], and Zhen Gao[1,†]

[1]State Key Laboratory of Optical Fiber and Cable Manufacturing Technology, Department of Electronic and Electrical Engineering, Guangdong Key Laboratory of Integrated Optoelectronics Intellisense, Southern University of Science and Technology, Shenzhen 518055, China.

[2]College of Physics and Electronic Engineering, Qujing Normal University, Qujing 655011, China

*These authors contributed equally to this work.

†Corresponding Email: xiatianzhi@mail.qjnu.edu.cn, xiahr@sustech.edu.cn, gaoz@sustech.edu.cn

## Abstract

Recent advances in topological physics have revealed a counterintuitive class of antichiral edge and surface states that propagate in the same direction along spatially separated parallel boundaries. To date, however, experimental realizations of antichiral states have been restricted to first-order topological phases, while their higher-order counterparts—antichiral hinge states—have remained experimentally elusive. Here, we report the first experimental observation of antichiral hinge states in a gyromagnetic photonic crystal that realizes a three-dimensional (3D) modified Haldane model with dimerized interlayer coupling. Through microwave near-field mapping, we directly resolve their defining signatures: nonreciprocal, co-propagating transport along four parallel hinges and characteristically tilted hinge-state dispersions. These results extend antichiral topology into the higher-order regime and provide a new platform for 3D nonreciprocal topological photonic devices.

## Introduction

The exploration and realization of novel topological phases of matter have become central themes in modern physics (1, 2), with far-reaching implications for both fundamental science and technological applications. First identified in condensed-matter systems, topological phases have since been extended to a wide range of physical platforms, including photonics (3), acoustics (4), mechanics (5), electrical circuits (6), thermotics (7), and ultracold atoms (8). A prominent hallmark of many such phases is the emergence of robust topological boundary states with unidirectional transport. A paradigmatic example is the quantum anomalous Hall effect (9, 10) in a Chern insulator (11, 12), which supports chiral edge states that propagate in opposite directions along parallel edges, as illustrated in Fig. 1a. Recent theoretical (13, 14) and experimental (15) advances have extended this concept to antichiral edge states, which propagate in the same direction along two parallel boundaries, as illustrated in Fig. 1b, thereby substantially expanding the possibilities for topologically protected transport. More recently, the concept of antichirality has been generalized from one-dimensional (1D) edge states to two-dimensional (2D) surface states (16, 17). To date, however, experimental demonstrations of antichiral topological states have been confined to first-order topological phases (15–17).

In parallel, the discovery of higher-order topological insulators (18, 19) has greatly expanded the classification of topological phases. In general, an $n$th-order topological phase in a $d$-dimensional system hosts robust $(d-n)$-dimensional boundary states ($n \geq 2$) protected by higher-order band topology. For example, a three-dimensional (3D) second-order topological insulator can support 1D chiral hinge states that propagate nonreciprocally in opposite directions along two pairs of parallel hinges (20–23), as illustrated in Fig. 1c. A natural yet unexplored frontier is therefore the integration of higher-order topology with antichiral propagation. Such a combination would not only give rise to new topological phases but also enable novel topological devices for robust multichannel wave transport. Although recent theoretical studies have proposed realizing antichiral hinge states by stacking two higher-order topological insulators with oppositely propagating chiral hinge states (24) or by employing a photonic synthetic dimension (25), their experimental realization in a single real-space higher-order topological structure, as illustrated in Fig. 1d, has remained elusive.

Here, we report the first experimental observation of antichiral hinge states in a 3D gyromagnetic photonic crystal (26–29). This system is the first photonic higher-order topological semimetal to exhibit the coexistence of nodal surfaces and nodal lines (30–36). The structure is formed by stacking 2D modified Haldane layers with dimerized interlayer couplings. It comprises gyromagnetic rods arranged in a 3D honeycomb lattice, with the two triangular sublattices magnetized in opposite directions. Microwave near-field mapping directly reveals copropagating antichiral hinge states along four parallel hinges, characterized by a tilted, fourfold-degenerate hinge dispersion with positive group velocity. Our work extends antichirality into the realm of higher-order topology for the first time and establishes a new route toward 3D nonreciprocal photonic devices that support robust, high-capacity transport through multiple spatial channels.

## Results

### Theoretical model

We begin with a 3D tight-binding model constructed by stacking 2D modified Haldane layers with dimerized interlayer couplings, as illustrated in Fig. 2a. The momentum-space Hamiltonian takes the block-diagonal form

$$H = H_{xy} \otimes \tau_0 + \sigma_0 \otimes H_z, \quad (1)$$

where $\sigma_0$ and $\tau_0$ are identity matrices acting on the sublattice $(A,\ B)$ and layer degrees of freedom, respectively. The in-plane Hamiltonian $H_{xy}$ describes the modified Haldane model and incorporates a symmetry-breaking tilt term essential for antichiral transport. The out-of-plane Hamiltonian $H_z$ introduces Su – Schrieffer – Heeger (SSH)-type dimerization of the interlayer couplings, thereby generating higher-order band topology (see Supplementary Material, Sec. 1). Because the two terms act on independent degrees of freedom, the eigenenergies and eigenstates of the full Hamiltonian can be written as $E_{n,m} = E_{xy,n} + E_{z,m}$ and $|\psi_{n,m}\rangle = |\psi_{xy,n}\rangle \otimes |\psi_{z,m}\rangle$, respectively. This factorization provides an intuitive mechanism for the formation of antichiral hinge states: they arise from the direct product of in-plane antichiral edge states, $|\psi_{xy,n}\rangle$, and out-of-plane SSH-type topological end states, $|\psi_{z,m}\rangle$ (37 – 39).

To elucidate the physical origin of these antichiral hinge states, we analyze the model's band structure and topological properties. The 3D Brillouin zone (BZ) is shown in Fig. 2b. The calculated bulk band structure in Fig. 2c reveals a distinctive topological semimetal phase characterized by the coexistence of frequency-shifted nodal lines (NLs; purple) and nodal surfaces (NSs; cyan). The NSs arise from crossings between the second and third bands, whereas the NLs originate from degeneracies between the first and second bands and between the third and fourth bands. To fully visualize these unique features, we plot the 3D momentum-space distribution of the band degeneracies in Fig. 2d. The corresponding iso-frequency contours obtained from local-density-of-states (LDOS) calculations are presented in Supplementary Material, Sec. 2. The in-plane $C_3$ symmetry protects the NLs located at the $K$ and $K'$ points, which extend continuously along the $k_z$ direction (purple line). Notably, these NLs are not isolated; instead, they are enclosed by NSs that form closed manifolds in 3D momentum space (colored surface). This distinctive nodal-surface-wrapped nodal-line configuration originates from the interplay between the in-plane Dirac dispersion and the out-of-plane nodal conditions imposed by the dimerized interlayer couplings (see details in Supplementary Material, Sec. 3). The projected band dispersion of a 3D modified Haldane model with dimerized interlayer couplings—finite in the *y*- and *z*-directions and periodic in the *x*-direction—is presented in Fig. 2e, from which we can see that the tilted four-fold-degenerate hinge dispersions (red line) connect the projections of the frequency-shifted NSs (cyan regions), whereas the tilted surface dispersions (green lines) emanate from the frequency-shifted NLs. Their copropagation along parallel boundaries confirms their antichiral character. Note that topological surface states also emerge on the top and bottom boundaries. However, since these states are largely hybridized with the bulk continuum, we omit them in the main text (see Supplementary Material, Sec. 4 for the detailed projected dispersion).

To establish the topological origin of the antichiral hinge states (40), we use the nested Wilson-loop method to calculate the nested polarization $p_y^{v_z}$ throughout the 3D momentum space (see Supplementary Material, Sec. 5 for the detailed calculation). As shown in Fig. 2f, $p_y^{v_z}$ is quantized to $1/2$ within the momentum interval bounded by the inner projected boundaries of the NSs. This half-quantized polarization is a hallmark of higher-order topology and underlies the bulk – hinge correspondence in our system. Notably, the nontrivial momentum region identified by this quantized 3D invariant coincides precisely with the momentum interval over which the antichiral hinge states appear. This agreement demonstrates that the localized, co-propagating antichiral hinge modes are protected by the 3D higher-order band topology (see Supplementary Material, Sec. 6 for the theoretical calculation with larger out-of-plane couplings).

**Design of a 3D gyromagnetic photonic crystal**

We then implement the theoretical model in a 3D gyromagnetic photonic crystal. As shown in the left panel of Fig. 3a, the unit cell consists of two types of gyromagnetic rods (red and blue) sandwiched between two pairs of permanent magnets (grey) for precise on-site magnetization control, together with two perforated metallic plates (yellow). To break time-reversal symmetry (TRS) while preserving the required sublattice symmetry, the permanent magnets generate a staggered magnetic field (white arrows), magnetizing the gyromagnetic rods in sublattices A and B along the $+z$ and $-z$ directions, respectively. The higher-order topology is achieved by dimerizing interlayer couplings through tailoring the geometry of the two perforated metallic plates, as illustrated in the right panel of Fig. 3a. The simulated bulk band structure of the 3D gyromagnetic photonic crystal [Fig. 3b] closely matches the theoretical results presented in Fig. 2c, further verifying the coexistence of NLs and NSs. The projected dispersion along the $k_x$ direction [Fig. 3c] reveals four degenerate antichiral hinge states (red lines) located within the bandgap of the antichiral surface states (green lines). Consistent with the tight-binding analysis, the antichiral hinge states occupy the momentum region bounded by the projections of the NSs, whereas the antichiral surface states connect the projections of the NLs. Both states exhibit characteristic tilted dispersions with positive group velocities, a hallmark of antichiral propagation. The simulated eigenmode profiles of the fourfold-degenerate antichiral hinge states at 8.1 GHz, presented in Fig. 3d, demonstrate that electromagnetic waves are tightly localized at the four parallel hinges (see Supplementary Material, Sec. 7, for the eigenfield profiles of the antichiral surface states). To further characterize their transport properties, four point sources (cyan stars) are placed at the hinge centers of a finite 3D gyromagnetic photonic crystal to excite these states. As shown in Fig. 3e, the excited electromagnetic waves remain confined to the hinges and propagate nonreciprocally in the same direction along all four parallel hinges, directly demonstrating the realization of antichiral hinge states.

**Experimental observation of antichiral hinge states**

Next, we experimentally demonstrate the antichiral hinge states in the fabricated 3D gyromagnetic photonic crystal. As shown in the upper panel of Fig. 4a, the sample consists of 20 × 4 × 5 unit cells along the $x$-, $y$-, and $z$-directions, respectively. The enlarged view in the lower panel of Fig. 4a illustrates the perforated copper plates and air foams (white), which accommodate the gyromagnetic rods and permanent magnets. Opposite magnetization directions on different sublattice sites are denoted by red and blue colors. Each gyromagnetic rod is sandwiched between a pair of permanent magnets with the corresponding magnetization orientation and fixed within the air foam. To characterize the nonreciprocal transport of the antichiral hinge states, we use a pair of source and probe antennas to measure the forward ($S_{21}$) and backward ($S_{12}$) transmission spectra along the two top hinges (red and blue lines in Fig. 4a). As shown in Figs. 4b and 4c, pronounced nonreciprocal transmission is observed within the frequency range of the antichiral hinge states (orange-shaded region), where the forward transmission (blue curves) substantially exceeds the backward transmission (red curves) for both hinges. These measurements directly reveal the co-propagating nature of antichiral hinge states along parallel hinges. To visualize the spatial profiles of these states, we place a point source (cyan star) at the center of each hinge and employ a probe antenna mounted on a robotic arm to sequentially sample the electric fields inside the air holes. As shown in Figs. 4d and 4e, the measured electric field distributions demonstrate unidirectional propagation of the antichiral hinge states along the $+x$ direction on both top parallel hinges (see Supplementary Material, Sec. 8, for the corresponding measurement results of the two bottom parallel hinges). By Fourier transforming the measured complex electric-field distributions from real to momentum space, we reconstruct the dispersions of the antichiral hinge states along the $k_x$ direction, as shown in Figs. 4f and 4g. The experimentally obtained dispersions (colormaps) agree well with the simulated results (red lines), with both hinges exhibiting tilted dispersions and positive group velocities, characteristic of antichiral propagation.

**Experimental demonstration of the robustness of antichiral hinge states**

Finally, we investigate the robustness of the antichiral hinge states. We first designed and fabricated a 3D gyromagnetic photonic crystal incorporating a sharp corner, as shown in Fig. 5a. Perfect-electric-conductor boundary conditions were imposed on two surfaces of the sample, whereas an absorbing boundary condition was applied to the third (see the experimental setup in Supplementary Material, Sec. 9). The electric-field distributions obtained from full-wave simulations (Fig. 5b) and microwave near-field measurements (Fig. 5c) show that the antichiral hinge states smoothly navigate the sharp corner without discernible backscattering. We further introduced a lattice defect along their propagation path by removing a YIG rod together with its pair of permanent magnets, as shown in Fig. 5d. Both the simulated (Fig. 5e) and measured (Fig. 5f) electric-field profiles demonstrate that the hinge states circumvent the defect with negligible attenuation. The measured transmission spectra and dispersions in the presence of the structural defect are provided in Supplementary Material, Sec. 10. Together, these results confirm the topologically protected robustness of the antichiral hinge states.

**Discussion**

In conclusion, we report the first experimental observation of antichiral hinge states in a 3D gyromagnetic photonic crystal, extending the concept of antichirality to higher-order topological phases. Beyond realizing antichiral hinge transport, we establish a new class of photonic higher-order topological semimetals featuring the coexistence of NLs and NSs in an unconventional NS-wrapping-NL configuration, together with antichiral surface and hinge states. As a 3D form of topologically protected one-way electromagnetic transport, antichiral hinge states offer new opportunities for the development of nonreciprocal photonic devices (see details in the Supplementary Material, Sec. 11). Our work broadens the scope of antichiral and higher-order topological phenomena and provides a versatile platform for exploring exotic states in magnetic systems with broken TRS.

## Methods

The numerical results in this study were obtained through simulations using the RF module of COMSOL Multiphysics. Both the copper plates and the permanent magnets are considered perfect electric conductors. The gyromagnetic material is yttrium iron garnet (YIG) ferrite, with relative permittivity 14.3. The relative permeability tensor of the gyromagnetic materials has the form:

$$[\mu_r] = \begin{bmatrix} \mu_m & \pm j\mu_k & 0 \\ \mp j\mu_k & \mu_m & 0 \\ 0 & 0 & 1 \end{bmatrix},$$

where $\mu_m = 1 + \omega_m(\omega_0 + i\alpha\omega)/[(\omega_0 + i\alpha\omega)^2 - \omega^2]$, $\mu_k = \omega_m\omega/[(\omega_0 + i\alpha\omega)^2 - \omega^2]$, $\omega_m = \gamma\mu_0 M_s$, $\omega_0 = \gamma\mu_0 H_0$, and $\mu_0 H_0$ is the external magnetic field (about 0.115 Tesla) along the z-direction, $\gamma = 1.759 \times 10^{11}\ C/kg$ is the gyromagnetic ratio, and $\omega$ is the operating frequency. The saturation magnetization is $M_s$ = 1780 Gauss, and $\alpha$ = 0.0088 is the damping coefficient.

In the experimental measurements, two microwave dipole antennas, functioning as the source and probe, are connected to a vector network analyzer (Keysight E5080). The complex electric field distributions are mapped by inserting the probe into the air holes one by one and scanning along the z direction in small steps. Then, the Fourier transformation was applied to the measured complex electric field distributions at each frequency to obtain the measured hinge dispersion relations.

**Data availability:** All data that support the plots within this paper and other findings of this study are available from the corresponding authors upon reasonable request.

**Code availability:** We use commercial software COMSOL Multiphysics to perform electromagnetic numerical simulations. Requests for computation details can be addressed to the corresponding authors.

**Acknowledgement**

Z.G. acknowledges funding from the National Key R&D Program of China (grant no.

2025YFA1412300), National Natural Science Foundation of China (grant no. 62361166627 and 62375118), Guangdong Basic and Applied Basic Research Foundation (grant no. 2024A1515012770), Shenzhen Science and Technology Innovation Commission (grants no. 20230802205352003), and High-level Special Funds (grant no. G03034K004). Z.W. acknowledges the funding from the National Natural Science Foundation of China (Grant No. 125B2093).

**Author contributions:**
Z.G. initiated the project. Z.G., T.Z.X., and H.R.X. conceived the idea and supervised the project. Z.Y.W. and T.Z.X. performed the numerical calculations. Z.Y.W. and H.R.X. performed the simulations. Z.Y.W. and T.Z.X designed the experiments. Z.Y.W. fabricated samples and carried out the measurements. Z.Y.W., T.Z.X., and H.R.X. analyzed data. Z.Y.W. drafted, and T.Z.X, H.R.X., and Z.G. revised the manuscript.

**Competing interests:** The authors declare no competing interests.

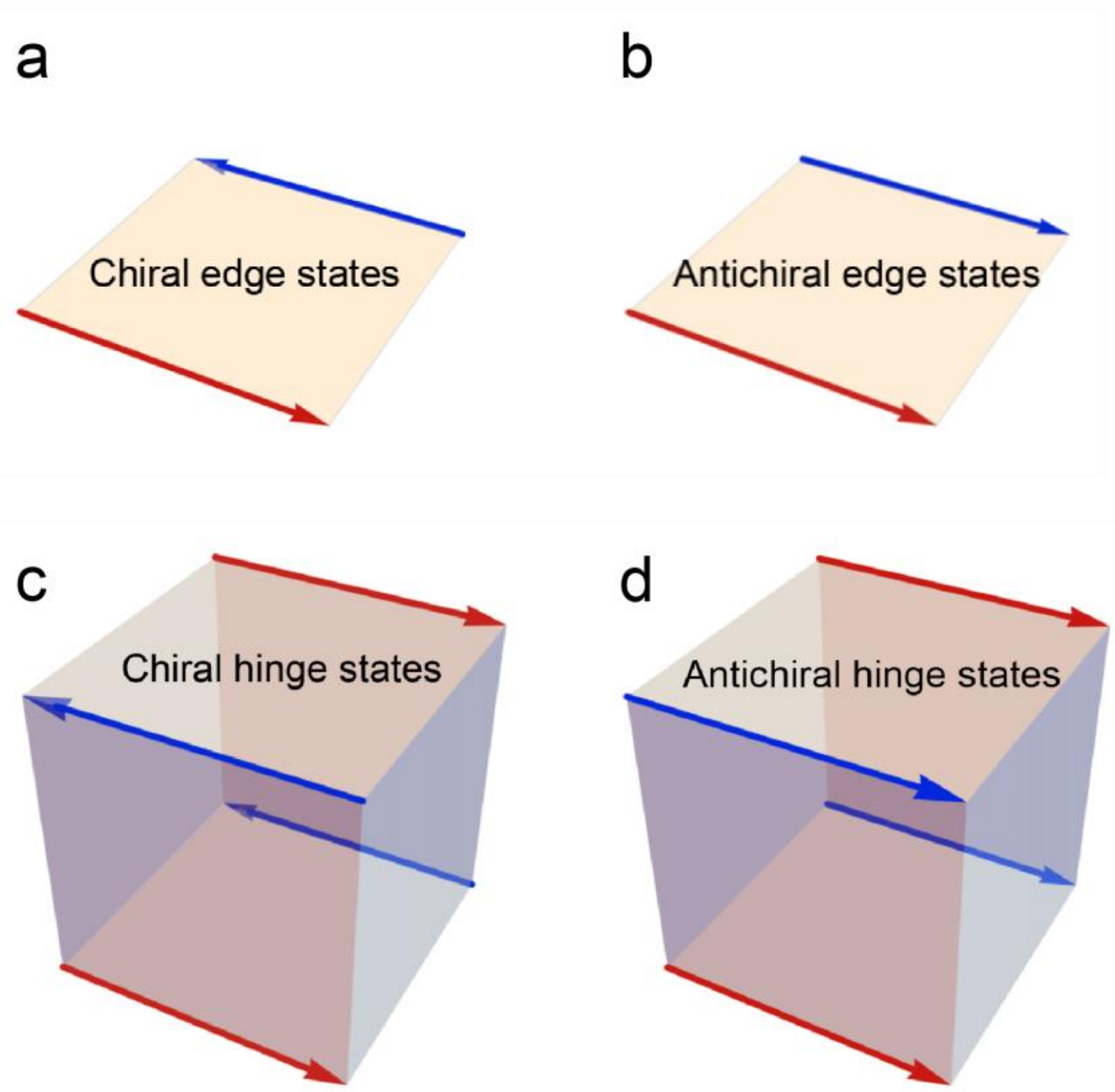

**Fig. 1 | Schematic illustrations of chiral and antichiral topological states in 2D and 3D systems.** **a** Chiral edge states that propagate in opposite directions along two parallel edges. **b** Antichiral edge states that propagate in the same direction along two parallel edges. **c** Chiral hinge states that propagate in opposite directions along two pairs of parallel hinges. **d** Antichiral hinge states that propagate in the same direction along two pairs of parallel hinges.

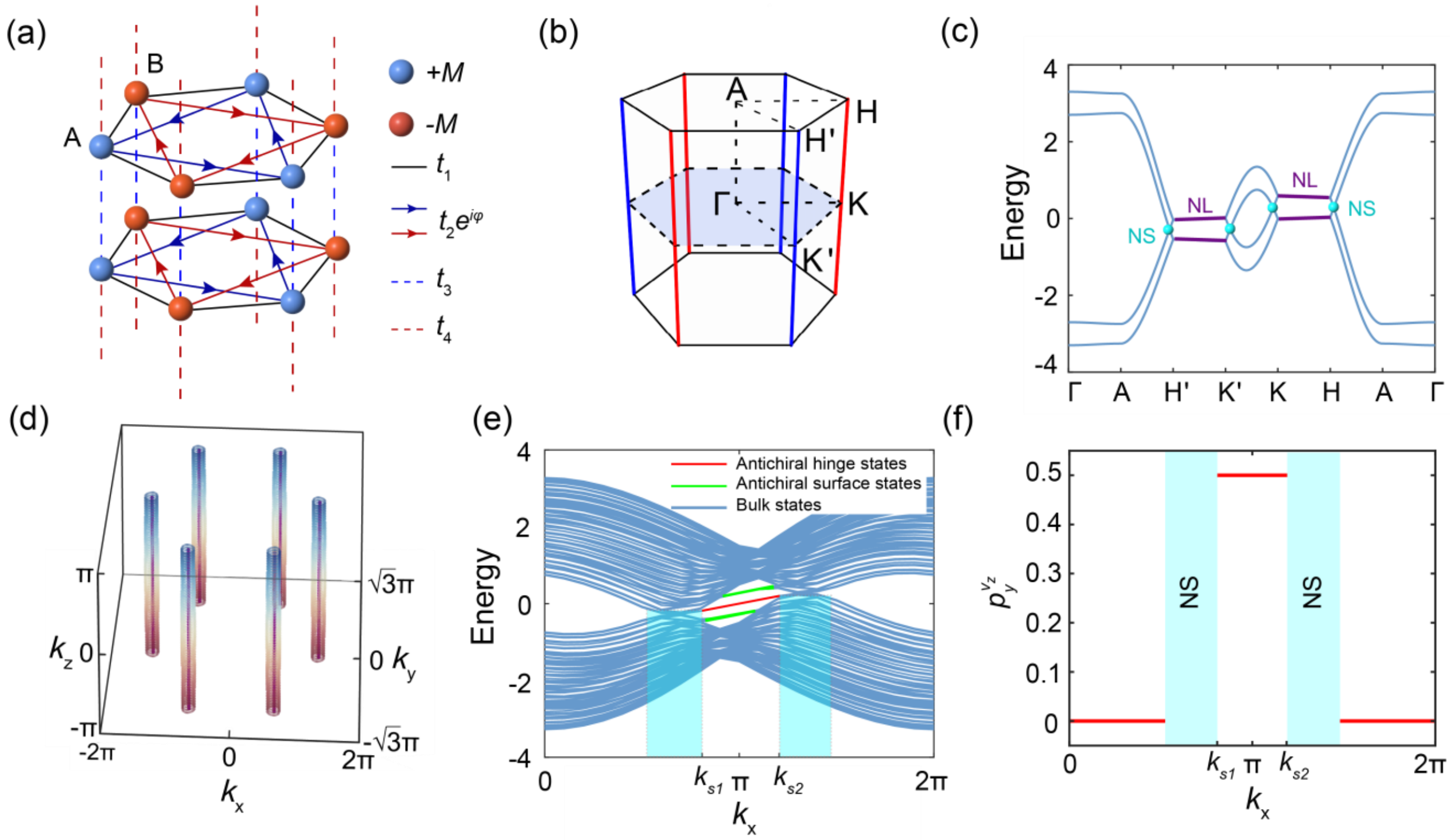

**Fig. 2 | Theoretical model. a** Schematic of a 3D modified Haldane model with dimerized interlayer couplings. Sublattices A and B are denoted by blue and red spheres, respectively. **b** The 3D BZ. **c** Calculated bulk band structure of the 3D tight-binding model with parameters $M = 0$, $t_1 = 1$, $t_2 = 1/18$, $\varphi = \pi/2$, $t_3 = 0.2t_2$, and $t_4 = 5t_2$. The cyan spheres represent the NSs, and the purple lines denote the NLs. **d** Visualization of the NSs (colored surfaces) and NLs (purple lines) in the 3D momentum space. **e** Projected band dispersions of a 3D modified Haldane model with dimerized interlayer coupling that is finite in the *y*- and *z*-directions and periodic in the *x*-direction. Red and green lines represent the antichiral hinge and surface states, respectively. The cyan-shaded region represents the NSs. **f** The nested polarization $p_y^{v_z}$ is evaluated along $k_x$ computed via nested Wilson loops for occupied bands .

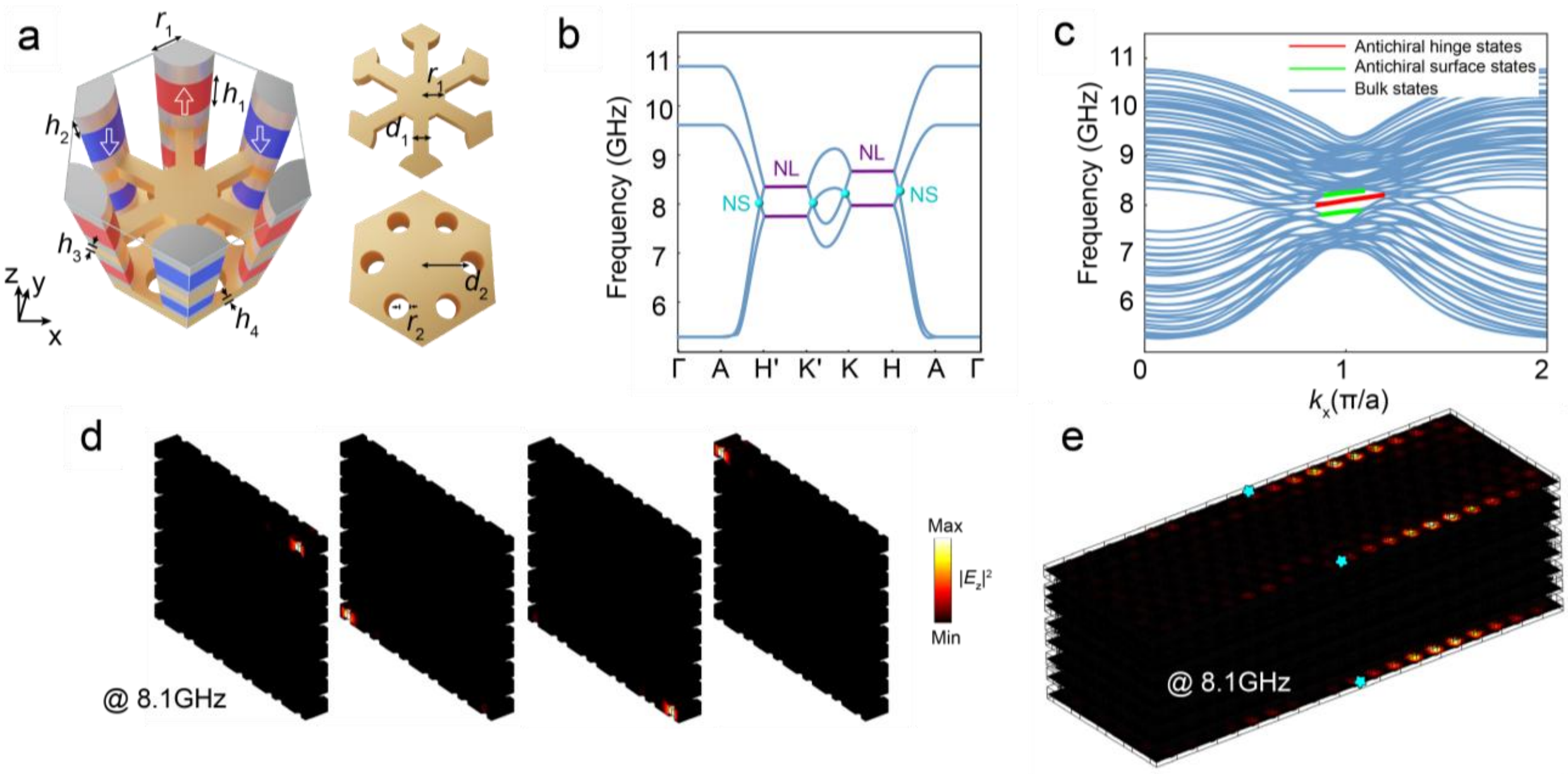

**Fig. 3 | Design of a 3D gyromagnetic photonic crystal. a** Left panel: unit cell of the 3D gyromagnetic photonic crystal. The lattice constants are $a = 7.5$ mm in the $xy$-plane and $a_z = 11$ mm along the $z$-direction. Other geometrical parameters are $r_1 = 1.3$ mm, $h_1 = 2$ mm, $h_2 = 1$ mm, $h_3 = 1$ mm, and $h_4 = 2$ mm, respectively. Right panel: top view of the two different perforated metallic plates, with geometric parameters $d_1 = 0.8$ mm, $d_2 = a/2.8$, $r_2 = r_1/2$, respectively. The white arrow indicates the direction of magnetization. **b** Simulated bulk band structure of the 3D gyromagnetic photonic crystal along high-symmetry lines. The cyan spheres represent the NSs, and the purple lines denote the NLs. **c** Simulated projected antichiral surface (green lines) and hinge (red lines) state dispersions along the $k_x$ direction. **d** Simulated electric field distributions of the eigenmode profiles of the fourfold-degenerate antichiral hinge states at $k_x = \pi/a$. **e** Simulated electric field distributions of the antichiral hinge states at 8.1 GHz, with the point sources (cyan stars) placed at the center of four parallel hinges. Scattering boundary conditions are applied to the left and right boundaries.

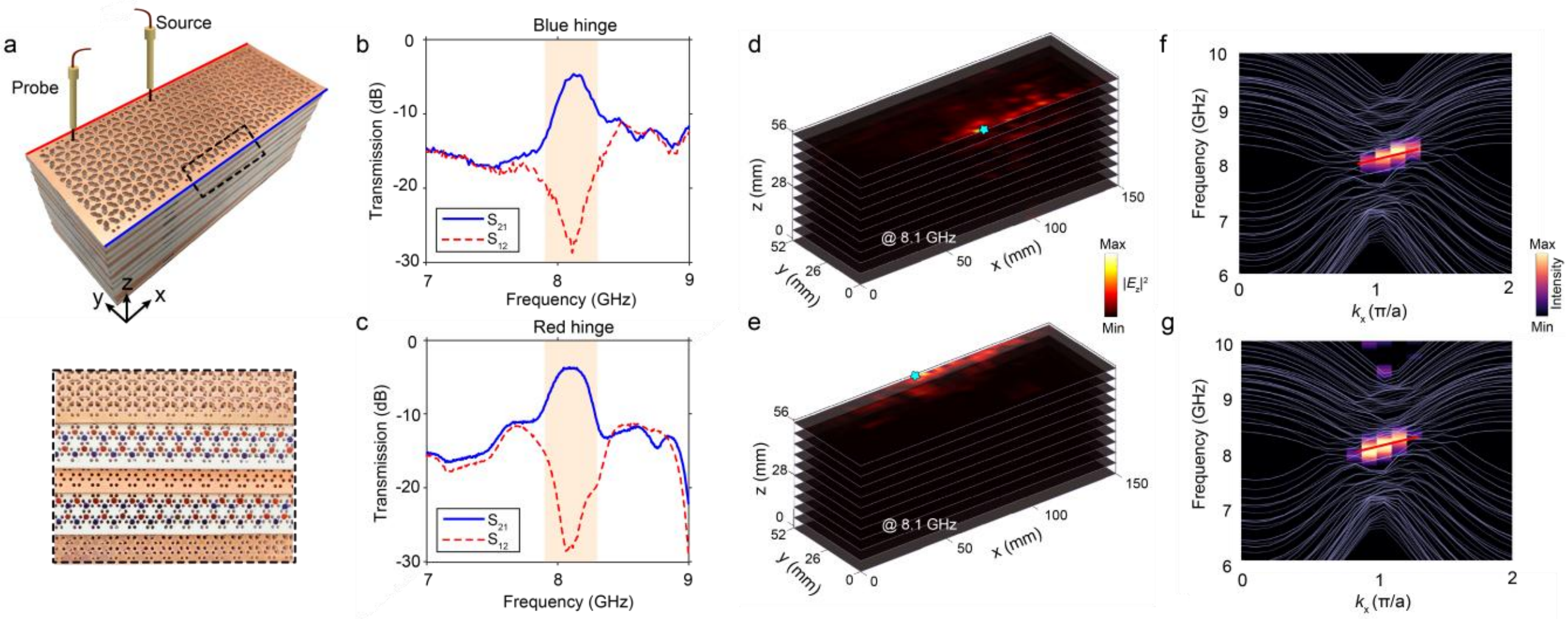

**Fig. 4 | Experimental observation of antichiral hinge states. a** Upper panel: Photograph of the fabricated 3D gyromagnetic photonic crystal that consists of $20 \times 4 \times 5$ unit cells in the *x*, *y*, and *z* directions, respectively. Lower panel: Enlarged photographs of the perforated copper plates and air foams (white), respectively. Red and blue disks denote the magnetized directions of gyromagnetic rods at different sublattice sites. **b, c** Measured forward ($\mathrm{S}_{21}$, blue line) and backward ($\mathrm{S}_{12}$, red line) transmission spectra at the two top hinges. The orange-shaded regions indicate the frequency range of the antichiral hinge states. **d, e** Measured electric field distributions of the antichiral hinge states at 8.1 GHz, with the source antenna (cyan star) positioned at the center of the hinges. **f, g** Measured (colormaps) and simulated (red lines) dispersions of the antichiral hinge states at the two top hinges along the $k_x$ direction.

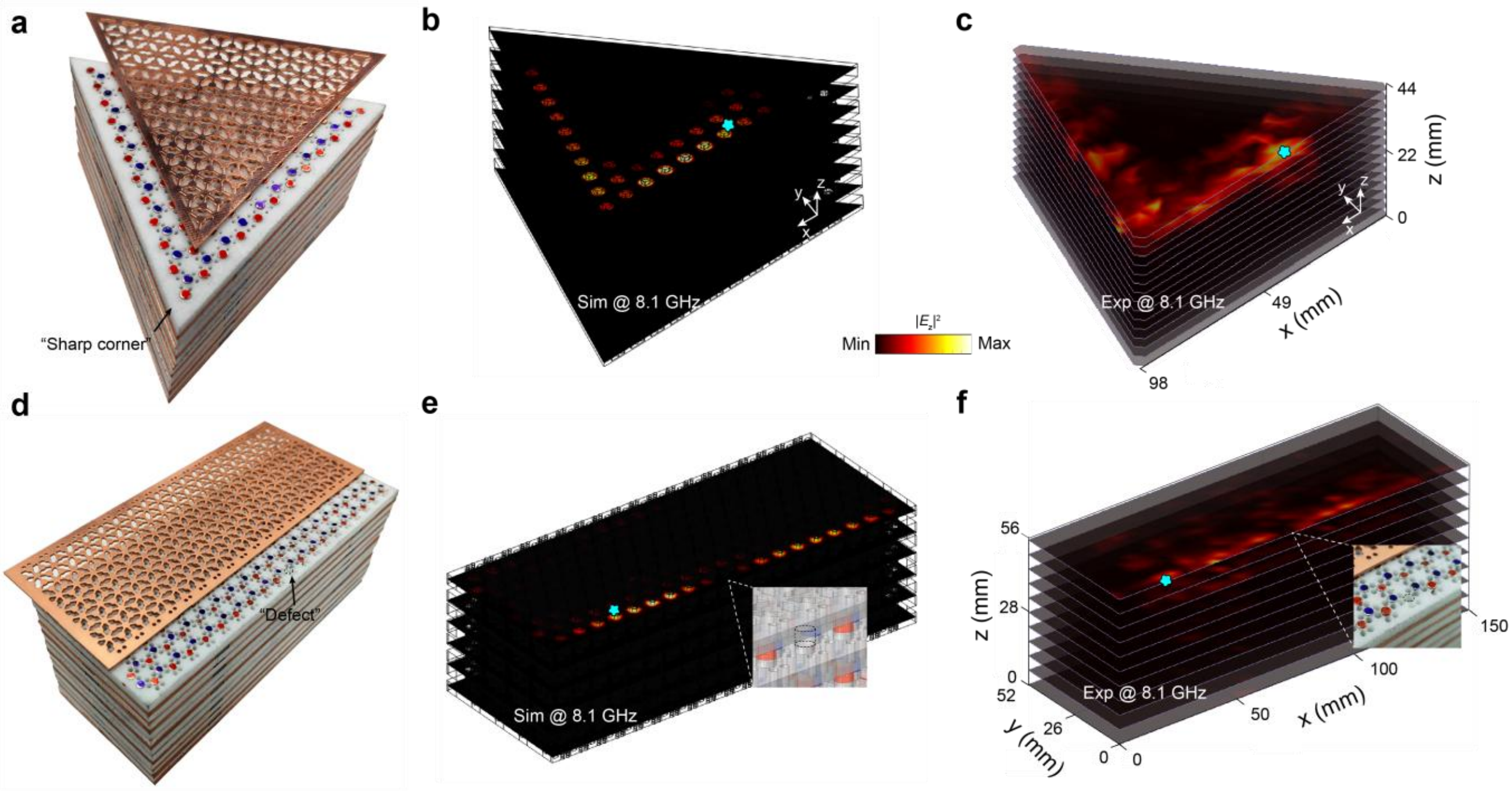

**Fig. 5 | Experimental demonstration of the robustness of antichiral hinge states against sharp corners and structural defects. a** Photograph of a fabricated 3D gyromagnetic photonic crystal with sharp corners. **b, c** Simulated (**b**) and measured (**c**) electric field distributions of the antichiral hinge states route around a sharp corner smoothly without backscattering. **d** Photograph of a fabricated 3D gyromagnetic photonic crystal containing a lattice defect in the top hinge. **e, f** Simulated (**e**) and measured (**f**) electric field distributions of the antichiral hinge states propagate through a lattice defect without backscattering.